# Electrically tunable quantum interference of atomic spins on surfaces

Hao Wang[1,2,†], Jing Chen[1,2,†], Peng Fan[1,2,†], Yelko del Castillo[3], Alejandro Ferrón[4], Lili Jiang[2], Zilong Wu[1,2],

Shijie Li[1,2], Hong-Jun Gao[1,2], Heng Fan[1,2], Joaquín Fernández-Rossier[3]*, Kai Yang[1,2]*

[1]Beijing National Laboratory for Condensed Matter Physics and Institute of Physics, Chinese Academy of Sciences, Beijing 100190, China

[2]School of Physical Sciences, University of Chinese Academy of Sciences, Beijing 100049, China

[3]International Iberian Nanotechnology Laboratory (INL), Avenida Mestre José Veiga, 4715-310 Braga, Portugal

[4]Instituto de Modelado e Innovación Tecnológica (CONICET-UNNE), and Facultad de Ciencias Exactas, Naturales y Agrimensura, Universidad Nacional del Nordeste, Avenida Libertad 5400, W3404AAS Corrientes, Argentina

*Corresponding authors: K.Y. (kaiyang@iphy.ac.cn); J.F-R. (joaquin.fernandez-rossier@inl.int)

[†]These authors contributed equally to this work.

**Controlling quantum interference near avoided energy-level crossings is crucial for fast and reliable coherent manipulation in quantum information processing. However, achieving tunable quantum interference in atomically-precise engineered structures remains challenging. Here, we demonstrate electrical control of quantum interference using atomic spins on an insulating film in a scanning tunneling microscope. Using bias voltages applied across the tunnel junction, we modulate the atomically-confined magnetic interaction between the probe tip and surface atoms with a strong electric field, and drive the spin state rapidly through the energy-level anticrossing. This all-electrical manipulation allows us to achieve Landau-Zener-Stückelberg-Majorana (LZSM) interferometry on both single spins and pairs of interacting spins. The LZSM pattern exhibits multiphoton resonances, and its asymmetry suggests that the spin dynamics is influenced by spin-transfer torque of tunneling electrons. Multi-level LZSM spectra measured on coupled spins with tunable interactions show distinct interference patterns depending on their many-body energy landscapes. These results open new avenues for all-electrical quantum manipulation in spin-based quantum processors in the strongly driven regime.**

Quantum interference, a fundamental manifestation of the wave-like nature of quantum particles, occurs when a system is prepared in a linear superposition of two states with a well-controlled relative phase. A paradigmatic example is Landau-Zener-Stückelberg-Majorana (LZSM) interference[1], which arises when a quantum two-level system is repeatedly driven through an anticrossing in the energy-level diagram, and undergoes multiple non-adiabatic transitions. The phases accumulated in each passage may result in constructive or destructive interferences. In the solid-state environment, LZSM interferometry has been harnessed for quantum control and qubit characterization in semiconductor quantum dots[2-4], superconducting qubits[5,6], and nitrogen-vacancy centers in diamond[7]. However, it remains a significant challenge to achieve tunable LZSM interference in an atomic-scale quantum architecture where multiple spins can be precisely assembled and controllably coupled[8-10].

Scanning tunneling microscope (STM) has been used to induce and probe quantum interference with atomic precision on solid surfaces, as demonstrated in quantum corrals[11], standing single molecules[12], and atomic-scale Josephson junctions[13,14]. In these systems, interference occurred in the orbital part of the wave function. By contrast, quantum interference based on spins on surfaces remains largely unexplored, despite their potential for quantum applications due to their longer coherence time, compared to charge states[15]. Recent development of electron spin resonance (ESR)–STM[16-19] has enabled the demonstration of Ramsey interference in individual surface spins[20]. Nevertheless, quantum interference of surface spins based on non-adiabatic transitions has not been reported so far—mostly owing to the challenge of precise spin manipulation to generate the required time-dependent Hamiltonian. Previous works on spin manipulation have primarily focused on controlling spin-state populations through scattering with tunnel current[17,21,22], but how to achieve rapid modulation of energy levels of surface spins remains elusive.

Here, we demonstrate electrical control of LZSM interferometry over the spin states on both individual Ti adatoms as well as Ti dimers on MgO films in a low-temperature STM. By tuning the magnetic interaction with the STM tip through the piezoelectric response of the surface atoms, we control their spin dynamics by applying time-dependent bias voltages ($V_{\text{bias}}$) (Fig. 1a), which modulate their energy levels and periodically drive their spin states through an anticrossing. The resulting LZSM interference of single Ti spins is magneto-resistively detected with a spin-polarized STM tip. The asymmetric interference pattern reveals the competition between a spin-transfer torque process and the energy-level modulation. We are also able to perform LZSM interferometry by modulating the frequency of radio-frequency (RF) voltages ($V_{\text{RF}}$) applied across the tunnel junction (Fig. 1a), providing an additional degree of freedom to engineer the spin quantum states. We further demonstrate multi-level LZSM interference in assembled Ti spin dimers with tunable magnetic interactions.

**Results**

**Electric control of energy detuning**

The Ti atoms were deposited on the two-monolayer MgO film grown on Ag(001), and are electrically accessible by measuring the time-averaged tunnel current[23]. The MgO serves as a decoupling layer which increases the coherence time of the Ti spins[23,24]. We focus on spin-1/2 Ti atoms adsorbed at the bridge site between two oxygen atoms of MgO[25]. The spin-polarized STM tip, prepared by transferring Fe atoms to the tip apex, is used to probe and control the Ti spins. Single-atom ESR signals are measured by applying a bias voltage $V_{\text{bias}}$ and an RF voltage $V_{\text{RF}}$ of frequency $\omega_{\text{RF}}$ across the STM tunnel junction[20,23].

The magnetic field experienced by a Ti spin (**S**) is the vector sum of the externally applied field $\mathbf{B}_{\text{ext}}$ and the effective field of the tip[26]. The tip field is modulated by $V_{\text{RF}}$, and thus has a static component $\mathbf{B}_{\text{tip}}$ and an oscillatory component $\Delta\mathbf{B}_{\text{tip}}\cos(\omega_{\text{RF}}t)$. The Hamiltonian is therefore given by[20,26]:

$$H = g\mu_B \mathbf{B}\cdot\mathbf{S} + g\mu_B \Delta\mathbf{B}_{\text{tip}}\cdot\mathbf{S}\cos(\omega_{\text{RF}}t) \qquad (1)$$

where $\mathbf{B} = \mathbf{B}_{\text{ext}} + \mathbf{B}_{\text{tip}}$ is the total static field, which sets the Zeeman splitting between the spin-up $|\uparrow\rangle$ and spin-down $|\downarrow\rangle$ states. The g-factor is about 1.8, and $\mu_B$ is the Bohr magneton[23,25].

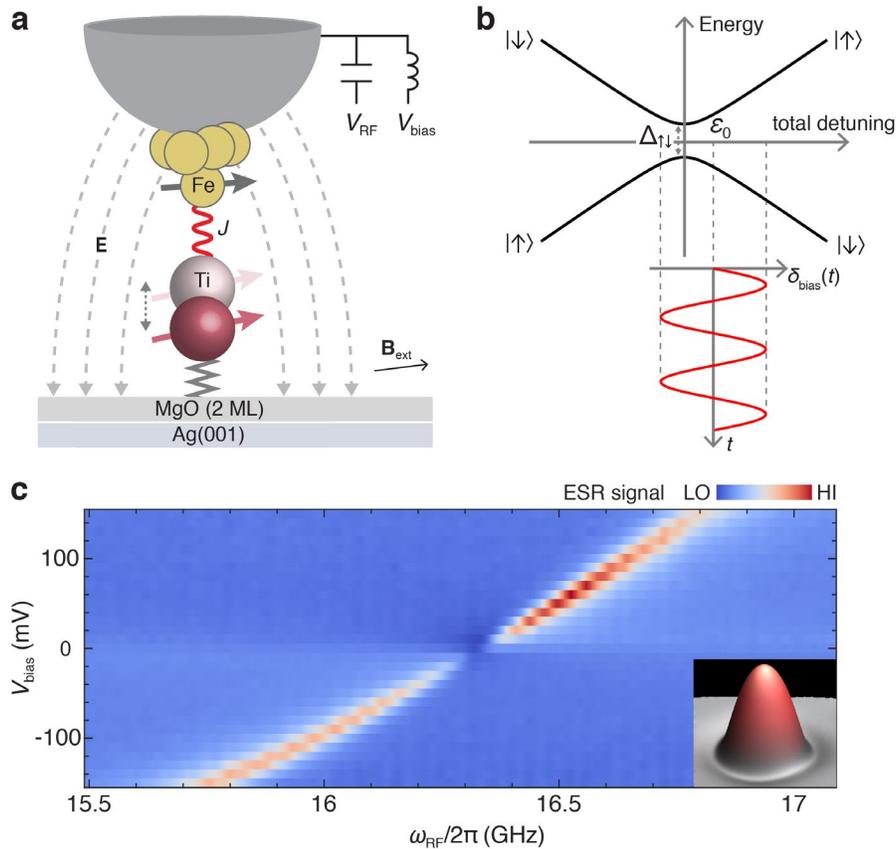

**Figure 1 | Electric control of the energy detuning of a single Ti spin. a,** Experimental setup showing a Ti atom on MgO and a spin-polarized STM tip with several Fe atoms attached to the tip apex. A modulated bias voltage $V_{\text{bias}}$ is applied across the tunnel junction, and the resulting AC component of electric field $E$ (dashed arrows) induces oscillation of the Ti atom (dashed double arrow), which alters the exchange interaction $J$ (indicated by the red curve) between the Ti and Fe atoms. The Ti-MgO bond is indicated by the zigzag curve. **b,** Energy-level diagram of a single Ti spin (driven by both $V_{\text{RF}}$ and a sinusoidally modulated $V_{\text{bias}}$) in the rotating frame as a function of total detuning, which is the sum of the static detuning $\varepsilon_0$ and modulated detuning $\delta_{\text{bias}}(t)$. The hybridization $\Delta_{\uparrow\downarrow}$ between $|\uparrow\rangle$ and $|\downarrow\rangle$ states is shown. **c,** ESR spectra of a single Ti spin as a function of RF frequency $\omega_{\text{RF}}$ and bias voltage $V_{\text{bias}}$, taken at a constant static tip-atom distance (setpoint: $V_{\text{sp}}$ = 50 mV, $I_{\text{sp}}$ = 60 pA; $B_{\text{ext}}$ = 0.51 T). The inset shows the STM image of the Ti atom.

Realizing LZSM interference requires two key ingredients[27]. One is the avoided level crossing, which can be obtained by making a transformation into a frame rotating with frequency $\omega_{RF}$. In the rotating frame, the oscillatory tip field becomes a static transverse magnetic field, which hybridizes the $|\uparrow\rangle$ and $|\downarrow\rangle$ states, opening an anticrossing of magnitude $\Delta_{\uparrow\downarrow} = \hbar\Omega_{Rabi}$ at zero energy detuning (Fig. 1b), where $\Omega_{Rabi}$ is the Rabi frequency.

The second ingredient is an adjustable energy detuning, which sets the energy difference between the $|\uparrow\rangle$ and $|\downarrow\rangle$ states. We controlled the detuning of the spin-1/2 Ti atom by the bias voltage $V_{bias}$, which induces a very strong electric field as high as ~1 GV/m across the STM junction due to the sub-nanometer tip-sample distance[18,26,28]. In turn, this results in a modulation of the spin splitting of the surface spin, which we probe using ESR-STM. Specifically, we measure the evolution of ESR spectra of a single Ti spin as a function of $V_{bias}$, taken at a constant static tip-atom distance (Fig. 1c)[18]. The ESR peak shifts almost linearly with $V_{bias}$. As the ESR frequency is proportional to $B_{ext} + B_{tip}$, this frequency shifts indicates that $B_{tip}$ increases monotonically with increasing $V_{bias}$, giving rise to an effective detuning shift of ~3 MHz/mV.

The spin-electric field coupling may arise from an atomic-scale piezoelectric effect, where the strong electric field alters the equilibrium position of the Ti atom on MgO by ~1 % of the Ti-MgO distance[18,26,28]. Since the spin interaction between the magnetic tip and the Ti atom depends exponentially on their distance[26], the piezoelectric displacement of the Ti atom results in a modified static $B_{tip}$ and thus the energy detuning (Fig. 1a). This electrical spin control of energy detuning offers architectural advantages for quantum spintronics because electric fields can be efficiently routed and confined in nanoscale circuits and adjusted faster compared to magnetic fields[2,18,29].

We can thus obtain a modulation $\delta_{bias}(t)$ of the energy detuning by adding a time-varying component to $V_{bias}$ (Fig. 1a). In the rotating frame, the spin Hamiltonian under $V_{RF}$ and modulated $V_{bias}$ is written as

$$H = [\varepsilon_0 + \delta_{bias}(t)]S_z + \Delta_{\uparrow\downarrow}S_x \qquad (2)$$

where the static detuning $\varepsilon_0 = \hbar(\omega_0 - \omega_{RF})$, and $\omega_0 = g\mu_B B/\hbar$ denotes the Larmor frequency. Here $z$-axis is the spin quantization axis as determined by the direction of the total static field **B**.

## LZSM interference of single spins

Our pulse sequence for LZSM measurement is illustrated in Fig. 2a. A bias voltage is applied at the tip-atom junction, with a sinusoidal modulation $V_{bias}(t) = V_{DC} + \delta V \sin(2\pi f t)$, on top of the RF voltage, with frequency $f \ll \omega_{RF}/2\pi$. Here $V_{DC}$ is the DC bias voltage; $\delta V$ and $f$ are the modulation amplitude and frequency, respectively. The modulated bias voltage $V_{bias}(t)$ drives the Ti spin repeatedly through the avoided level crossing (Fig. 1b), and we probe its effect on the steady-state spin occupation using single-atom ESR, which is sensitive to the change of spin population[20,23]. Figure 2b shows the LZSM pattern as a function of modulation frequency $f$ and static detuning $\varepsilon_0$ ($\propto \omega_{RF} - \omega_0$). For each static detuning, we measured the ESR signal at different modulation frequencies (Fig. 1b). The tunnel current shows clear LZSM interference fringes as a result of the phase accumulation between consecutive Landau-Zener transitions[1].

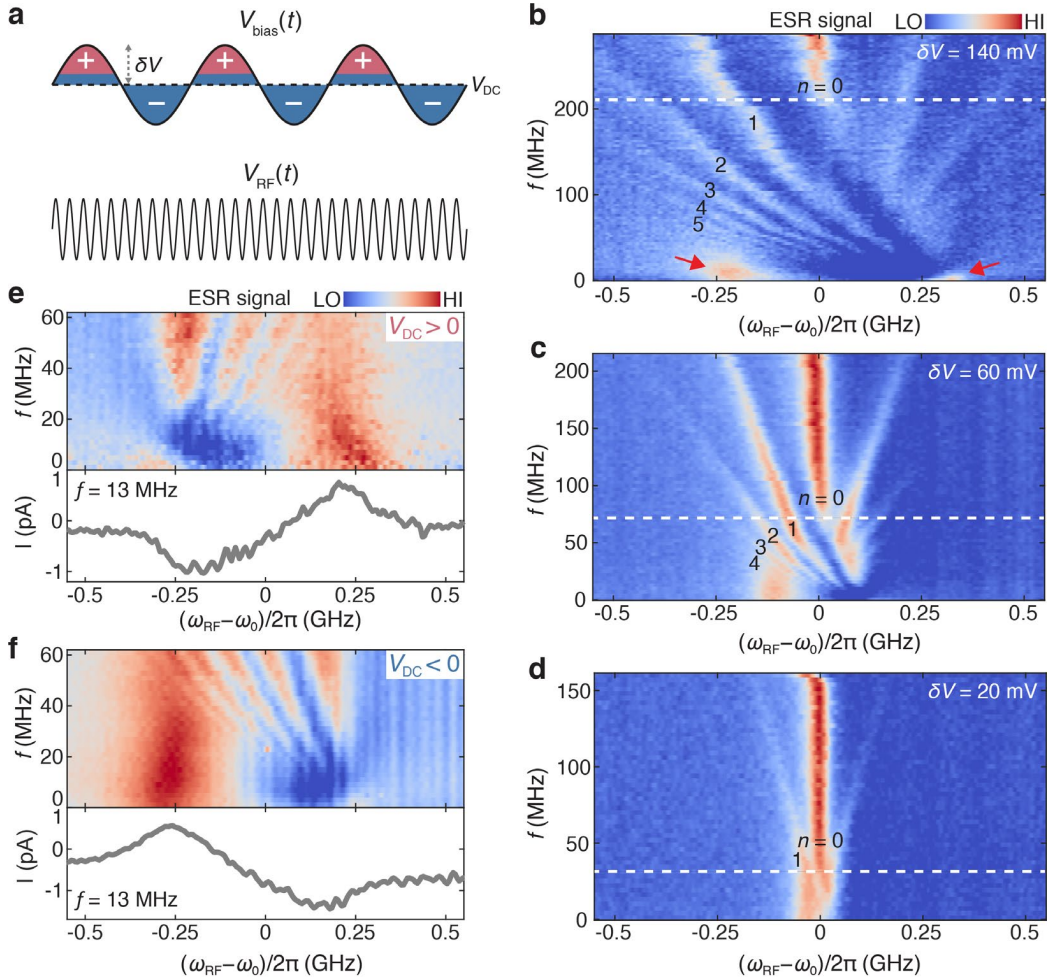

**Figure 2 | LZSM interference of a single Ti spin. a,** Schematics of the pulse sequences for the LZSM measurement. Both $V_{bias}$ and $V_{RF}$ are sinusoidal in time. $V_{DC}$ and $\delta V$ are the DC component and modulation amplitude of $V_{bias}$. The polarity of $V_{bias}$ is indicated by + (red regions) and − (blue regions) signs. **b-d,** ESR spectra as a function of detuning $\omega_{RF} - \omega_0$ and modulation frequency $f$, measured at a fixed modulation amplitude $\delta V$ of 140, 60 and 20 mV, respectively ($V_{DC}$ = −50 mV, $V_{RF}$ = 20 mV; $B_{ext}$ = 0.58 T; setpoint: $V_{sp}$ = 50 mV, $I_{sp}$ = 60 pA). The multiphoton resonances at $|\omega_{RF} - \omega_0|/2\pi = nf$ are labeled. Dashed white lines indicate the onset of motional averaging. Red arrows in **b** indicate the two main peaks at slow modulation. **e,f,** ESR spectra measured at low modulation frequencies $f$ with positive (**e**) and negative (**f**) DC voltages $V_{DC}$ ($V_{DC}$ = 50, −40 mV, $\delta V$ = 100, 80 mV, $V_{RF}$ = 15 mV; $B_{ext}$ = 0.52 T; setpoint: $V_{sp}$ = 50 mV, $I_{sp}$ = 50 pA). The lower panels show the ESR spectra at $f$ = 13 MHz.

The LZSM patterns in Fig. 2b exhibit various spectroscopic features depending on how fast the spin is driven, which according to our model (Supplementary Sec. 2), are governed by the dimensionless parameter $2\pi f T_2$. In the slow limit ($2\pi f T_2 \ll 1$), that corresponds to $f < 10$ MHz, the spectra exhibit two main ESR peaks (red arrows in Fig. 2b). This regime can be interpreted as if we were performing conventional ESR-STM experiments, but with a range of resonance frequencies following an arcsine distribution (Fig. 1b). In contrast, the interference pattern is observed in the coherent limit ($f > 20$ MHz), where consecutive traversals of anticrossing take place within the spin coherence time $T_2$. In this regime, a complex pattern with additional sidebands is observed. These ESR side peaks, appearing at $|\omega_{RF} - \omega_0|/2\pi = nf$, correspond to excitation processes driven by the adsorption of $n$ photons ($n \leq 5$, see Fig. 2b)[1]. The dressed spin is resonantly excited when the dressed energy splitting $\hbar|\omega_{RF} - \omega_0|$ matches the energy of $n$-photon (Supplementary Sec. 1). Note that the observation of the multiphoton process requires a very high driving field, which is easily fulfilled in the STM junction due to the sub-nanometer tip-atom distance.

Further increasing the modulation frequency $f$, a strong ESR peak appears at zero static detuning ($\omega_{RF} = \omega_0$), which corresponds to the motional average of the two main peaks at small modulation frequency[30].

We further studied the dependence of the LZSM spectra on the modulation amplitude $\delta V$ (Fig. 2b-d and see also Extended Data Fig. 1b-d). The threshold $f$ of the motional averaging increases with larger $\delta V$, as indicated by the dashed white lines in Fig. 2b-d, as shorter timescale is required to discriminate between the two main peaks reached by the energy detuning modulation. This is a demonstration of the energy-time uncertainty. In addition, as the modulation amplitude $\delta V$ increases, higher-order ESR side peaks become stronger, indicating higher-order photon modes increasingly participate in the excitation process, while the energy separation between adjacent photon-assisted modes is independent of $\delta V$.

We also measured the LZSM interference as a function of driving amplitudes $\delta V$ for a fixed modulation frequency $f$, and observed multiphoton resonances within a V-shaped region (Extended Data Fig. 1), as the driving amplitude needs to large enough to reach the avoided level crossing (Fig. 1b).

**Spin-transfer torque in LZSM interference**

As shown in Fig. 2b, the LZSM spectra exhibit pronounced asymmetries with respect to zero static detuning ($\omega_{RF} = \omega_0$), which cannot be captured by the conventional LZSM theory[1], and is in contrast to the symmetric patterns observed in other quantum systems[2-4].

This asymmetric pattern results from the spin-transfer torque on the Ti atom under the influence of the spin-polarized tunnel current[19,21,31]. The spin-transfer torque process is illustrated in Fig. 3a. At a positive $V_{bias}$, the inelastic tunneling electron ($\Delta\sigma = +1$) is able to cause a spin flip of the Ti atom ($\Delta m_{Ti} =$

−1) as the total spin angular momentum is conserved during the spin-scattering event. Reversing the polarity of $V_{\text{bias}}$ and thus the direction of tunnel current drives the Ti spin to the opposite direction ($\Delta m_{\text{Ti}} = +1$).

During the quantum interference, the atomic-scale spin-transfer torque process competes with the energy-level modulation (Fig. 3b). Since the center bias voltage $V_{\text{DC}}$ of $V_{\text{bias}}$ is nonzero, the amplitude of the spin-polarized current at positive $V_{\text{bias}}$ (indicated by + in Fig. 2a and 3b) differs from that at the negative $V_{\text{bias}}$ (indicated by − in Fig. 2a and 3b). This makes the symmetry of the LZSM pattern highly dependent on the polarity of $V_{\text{DC}}$, and the LZSM pattern is flipped with respect to $\omega_{\text{RF}} = \omega_0$ as the polarity of $V_{\text{DC}}$ is reversed (Extended Data Fig. 2).

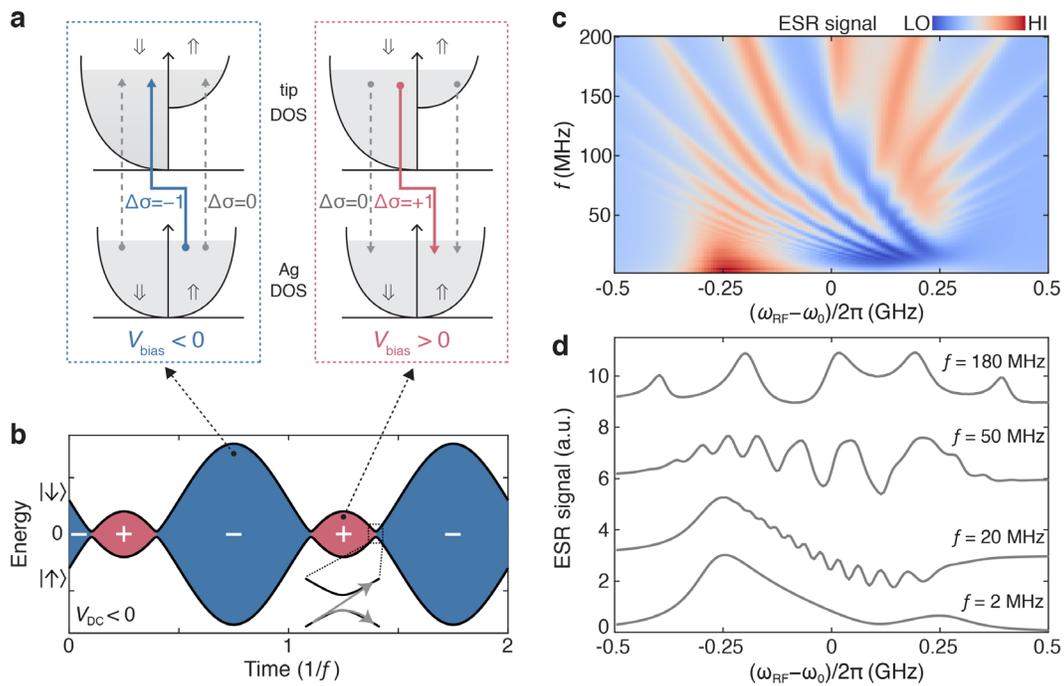

**Figure 3 | Simulation of the LZSM interference by considering the spin-transfer torque effects using the generalized Bloch equations. a,** Schematic of the spin-transfer torque on the Ti atom under the influence of spin-polarized tunnel current at different bias polarities, showing $\Delta\sigma = +1$ (red arrow), $\Delta\sigma = −1$ (blue arrow), and $\Delta\sigma = 0$ (dashed arrows) electron tunneling between the spin-dependent (⇑ or ⇓) densities of states (DOS) of Ag and STM tip. The bias is given with respect to the sample. **b,** Time evolution of energy levels of the Ti spin driven by bias voltage modulation for $V_{\text{DC}} < 0$. The polarity of $V_{\text{bias}}$ is indicated by + (red regions) and − (blue regions) signs. **c,** Simulated ESR signals as a function of detuning $\omega_{\text{RF}} - \omega_0$ and modulation frequency $f$ for a fixed modulation amplitude $\delta V$ of 140 mV. Simulations parameters: $\omega_0$ = 15.5 GHz, $\Gamma_{\text{STT}}(t) = 11 - 9\sin(2\pi ft)$ MHz, $\Delta_{\uparrow\downarrow}$ = 60 MHz, $\delta_{\text{bias}}(t) = 40\sin(2\pi ft)$ MHz, $V_{\text{DC}} = -50$ mV, $G_{\text{junc}}$ = 30 pS, $\alpha$ = 0.5, $\langle S_{\text{tip}}^z \rangle$ =1, $\langle S_{\text{tip}}^{xy} \rangle$ =0.5, $V_{\text{RF}}$ = 30 mV, $T_1$ = 400 ns, $T_2$ = 200 ns. See Supplementary Sec. 2 for the meaning of each parameter. **d,** Simulated ESR signals at $f$ = 2, 20, 50 and 180 MHz, respectively.

To quantitatively understand the asymmetric LZSM pattern, we calculate the time evolution of the periodically driven Ti spin using the generalized Bloch equations, and consider coherent driving, spin relaxation and decoherence, as well as the effect of spin-transfer torque by tunneling electrons during the energy-level modulation (Supplementary Sec. 2). The simulation results (Fig. 3c and Extended Data Fig. 3) agree well with the measured asymmetric LZSM pattern. As shown in Fig. 3c,d, the simulation reproduces the evolution from the two main peaks under very slow modulation, to the peak-dip spectra as well as asymmetric stripe patterns, and eventually to the more symmetric interference pattern at the high-$f$ regime.

The spectroscopic asymmetry is more pronounced in the low-$f$ regime ($f \sim 5 - 20$ MHz) of the LZSM spectra, where it manifests as a peak-dip lineshape in the ESR spectra (Fig. 2e,f and Fig. 3c,d). The dip occurs because the change in spin-state population, driven by the spin-transfer torque, cannot keep up with the rapid bias voltage modulation. This could occur for certain bias polarity during the modulation cycle, for example the positive bias cycle in Fig. 2f where the center bias voltage $V_{\text{DC}}$ is negative, leading to the dip in the spectra at $\omega_{\text{RF}} > \omega_0$. In contrast, for the negative bias cycle, spin-state population could still reach the nonequilibrium set by the larger negative spin-polarized current, resulting in the ESR peak.

When the bias voltage modulation is much faster than the spin relaxation, the change of spin-state population induced by spin-transfer torque becomes almost negligible over each cycle of the rapid energy-level modulation, leading to a less asymmetric interference pattern (Fig. 2b-d and Fig. 3c,d).

The asymmetric LZSM patterns highlights the dissipative action of the spin transfer-torque during quantum infereference[19]. The presence of the spin-polarized current also leads to a nonequilibrium initialization of the Ti spin, and thus could overcome the thermal population constrains.

**LZSM interferometry using frequency modulation**

In addition to modulating the bias voltage as shown above, we are also able to perform LZSM interferometry using frequency-modulated RF voltage[32]. The frequency modulation of $V_{\text{RF}}$ effectively modulates the fictitious field along the $z$-axis in the rotating frame, resulting in a tunable energy detuning. Compared to the bias voltage modulation, modulating the frequency of $V_{\text{RF}}$ offers the advantage of achieving much larger energy detuning, and also reduces spin scattering of Ti atom by tunnel current, leading to a longer $T_2$ time.

Specifically, we apply an RF voltage with modulated frequency $\omega_{\text{RF}}(t) = \omega_{\text{RF}} + \delta_{\text{RF}}(t)$ to the tunnel junction, where $\omega_{\text{RF}}$ is the center frequency and $\delta_{\text{RF}}(t)$ describes the frequency modulation. The corresponding spin Hamiltonian in the rotating frame is (Supplementary Sec. 3)

$$H = [\varepsilon_0 - \hbar\delta_{\text{RF}}(t)]S_z + \Delta_{\uparrow\downarrow}S_x \qquad (3)$$

Figure 4b shows the LZSM interference pattern measured with a sinusoidal frequency modulation: $\delta_{RF}(t) = \delta_{RF}\sin(2\pi f t)$, where $\delta_{RF}$ is the modulation amplitude (Fig. 4a). The frequency modulation results in a sinusoidal modulation of the energy detuning (Eq. 3). The reduced linewidth of the ESR peaks compared to Fig. 2b-d suggests an improved $T_2$ time due to a smaller averaged tunnel current flowing through the Ti atom. Importantly, a much larger energy detuning (~1 GHz) can be achieved compared to the bias voltage modulation, by simply increasing the frequency modulation amplitude $\delta_{RF}$. This combination of longer $T_2$ time and larger energy detuning enables the observation of higher-order multiphoton resonances with up to 8 photons. Furthermore, the LZSM spectra become symmetric since a constant bias voltage is applied, thereby eliminating the influence of spin-transfer torque modulation. This symmetric pattern is well reproduced by our simulation (Fig. 4c). We also measured LZSM interference as a function of $\delta_{RF}$ while keeping the modulation frequency $f$ fixed, and observed symmetric patterns in V-shaped regions (Extended Data Fig. 4).

In addition, using frequency modulation of $V_{RF}$, we can realize quantum interference under more complex situations. For instance, we realized the quantum interference in a two-level system under the modulation of both energy detuning and avoided anticrossing, using two frequency-modulated RF components (Fig. S2). This measurement provides new opportunities for quantum control, which is complementary to spin rotations using Rabi oscillations[10,20].

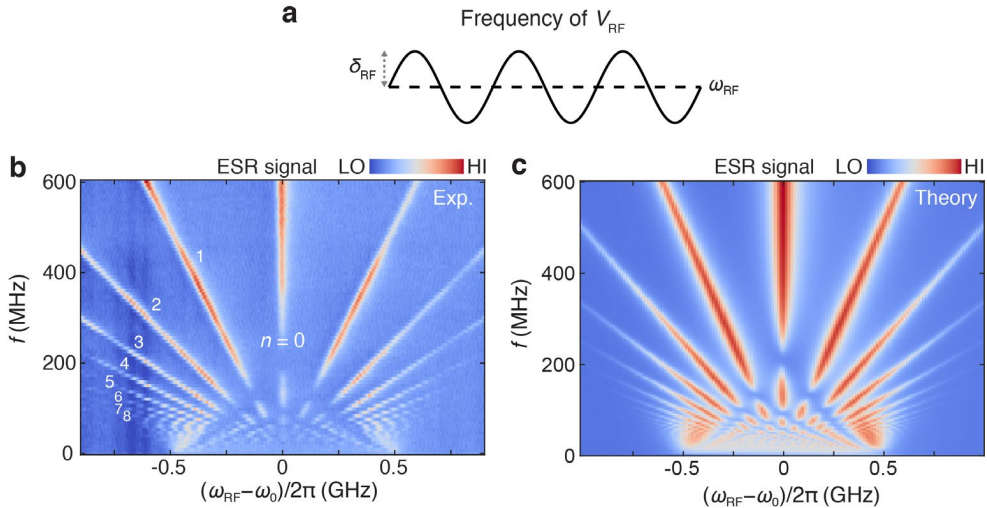

**Figure 4 | LZSM interference measured with frequency modulation of $V_{RF}$. a,** Schematics of the pulse sequences for the LZSM measurement in **b**. A sinusoidally frequency-modulated $V_{RF}$ is used, with a center frequency $\omega_{RF}$ and modulation amplitude $\delta_{RF}$. **b,** ESR spectra as a function of detuning $\omega_{RF} - \omega_0$ and modulation frequency $f$, measured using a sinusoidal frequency modulation of $V_{RF}$ ($\delta_{RF}$ = 0.5 GHz, $V_{RF}$ = 10 mV, $V_{bias}$ = 50 mV; $B_{ext}$ = 0.60 T; setpoint: $V_{sp}$ = 50 mV, $I_{sp}$ = 30 pA). **c,** Simulations of LZSM interference with frequency modulation of $V_{RF}$ using the

generalized Bloch equations. Simulations parameters: $\omega_0$ = 15.5 GHz, $V_{DC}$ = 50 mV, $\delta_{RF}$ = 500sin($2\pi f t$) MHz, $\Delta_{\uparrow\downarrow}$ = 60 MHz, $G_{junc}$ = 30 pS, $\alpha$ = 0.5, $\langle S_{tip}^z \rangle$ = 1, $\langle S_{tip}^{xy} \rangle$ = 0.5, $V_{RF}$ = 30 mV, $T_1$ = 400 ns, $T_2$ = 200 ns.

**LZSM interference of coupled spins**

We further demonstrate LZSM interference in multi-level systems using two coupled Ti spins with tunable interaction. The LZSM patterns reveal spectroscopic information about the energy level diagrams of spin dimers, including the occurrence of avoided level crossings as well as the magnetic interaction, and can be utilized to characterize the many-body energy levels of more complex quantum magnets[8,15,33].

The Ti spins are coupled by antiferromagnetic interaction ($J > 0$) (Fig. 5a), and the eigenstates consist of spin singlet state $|S\rangle$ and triplet states $|T_+\rangle$, $|T_0\rangle$ and $|T_-\rangle$ (refs. 17,18,23,25). Similar to the single spin, $V_{bias}$ can be used to control the energy detuning and thus the Zeeman energy of the spin $\mathbf{S}_1$ under the STM tip. As $V_{bias}$ varies, the energy-level diagram exhibits an avoided level crossing between $|S\rangle$ and $|T_0\rangle$ (Fig. 5b), which is detected by ESR spectra measured on $\mathbf{S}_1$ (Fig. 5c). Unlike the single spin (Fig. 1c), the ESR frequencies of coupled spins (I-IV) vary nonlinearly with $V_{bias}$ due to the spin coupling. In the following, we tune the two spins to the maximal level of entanglement by adjusting $V_{bias}$ to position the system at the anticrossing point (red dashed lines in Fig. 5b,c)[17,23]. At this point, the two spins experience the same Zeeman splitting and thus have the same Larmor frequency $\omega_0$.

We first drive the spin dimer with a frequency-modulated $V_{RF}$ applied on $\mathbf{S}_1$ (Fig. 5a). The corresponding spin Hamiltonian in the rotating frame defined by the transformation operator $e^{-i\omega_{RF} t (S_{1z}+S_{2z})}$, is written as (Supplementary Sec. 4):

$$H(t) = J\mathbf{S}_1 \cdot \mathbf{S}_2 + \hbar[\omega_0 - \omega_{RF} - \delta_{RF}(t)](S_{1z} + S_{2z}) + \Delta_{\uparrow\downarrow} S_{1x} \quad (4)$$

where $\omega_0$ is the Larmor frequency, $\delta_{RF}(t) = \delta_{RF} \sin(2\pi f t)$ denotes the frequency modulation of $V_{RF}$, and $\Delta_{\uparrow\downarrow}$ is proportional to the Rabi frequency of $\mathbf{S}_1$. Note that in the rotating frame, the frequency modulation affects the energy detuning of both spins, although the RF voltage is applied only on the spin $\mathbf{S}_1$. We plot the energy-level diagram as a function of static detuning $\hbar(\omega_{RF} - \omega_0)$ in Fig. 5d ($\delta_{RF} = 0$), which shows three avoided level crossings labelled as $\Delta_{S,-}$, $\Delta_{+,-}$ and $\Delta_{+,S}$. These anticrossings are opened due to $\Delta_{\uparrow\downarrow}$, and are separated by the coupling strength $J$ along the $x$-axis.

The resulting multi-level LZSM spectra as a function of modulation amplitude $\delta_{RF}$, encoding information about the energy-level spectrum, are displayed for three spin dimers with increasing coupling strength $J$ (Fig. 5e-g). When $\delta_{RF}$ is larger than $J/2$, two or three adjacent anticrossings can be traversed within a single modulation cycle (Fig. 5d), leading to the formation of diamond-like spectroscopic features in the spectra[6]. The boundaries of the diamonds indicate when the total detuning $\hbar(\omega_0 - \omega_{RF}) \pm \hbar\delta_{RF}$ reaches an anticrossing in the energy-level diagram. For example, in Fig. 5f, at the specific static detuning indicated by the vertical white line, three avoided crossings $\Delta_{+,-}$, $\Delta_{+,S}$ and $\Delta_{S,-}$ are sequentially reached with increasing $\delta_{RF}$, indicated by the three white arrows. Note that the first excited state is populated in the rotating frame, giving rise to the lower-half of the diamonds. In addition, the diagonal length of the diamonds corresponds to the coupling strength $J$, and thus the size of the diamonds increases with larger coupling (Fig. 5e-g). These results show that the LZSM pattern can be used to characterize the internal structures of the energy-level spectrum with multiple avoided level crossings.

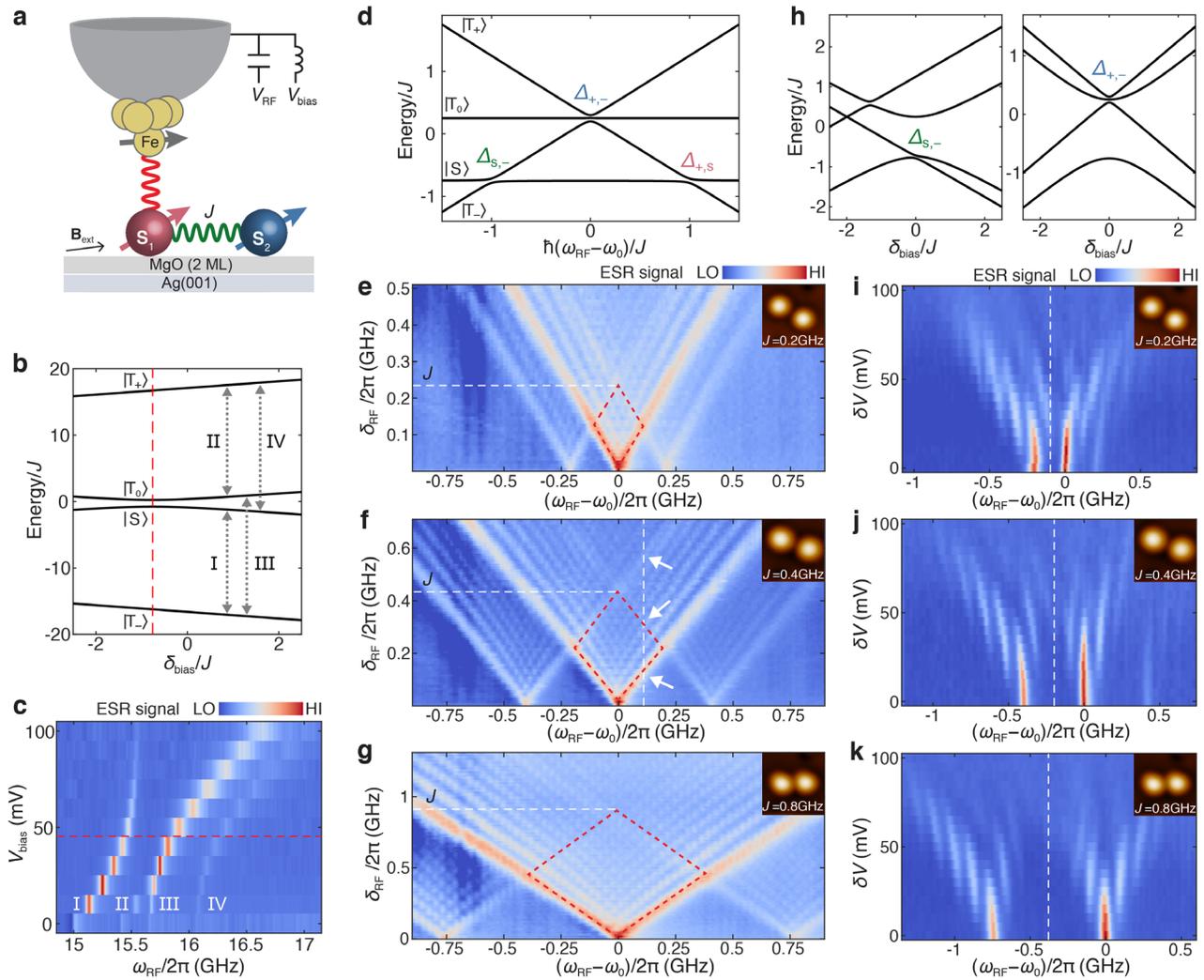

**Figure 5 | Tuning the LZSM interference of coupled spins. a,** Schematic showing two Ti spins with a coupling $J$ (green curve). The tip-Ti coupling is indicated by the red curve. **b,** Energy-level diagram of two spins as a function of the energy detuning $\delta_{bias}$ due to $V_{bias}$. $|S\rangle$ and $|T_0\rangle$ exhibit an anticrossing (red dashed line). **c,** ESR spectra of a Ti spin dimer ($J$ = 0.4 GHz) as a function of $\omega_{RF}$ and $V_{bias}$ (setpoint: $V_{sp}$ = 50 mV, $I_{sp}$ = 150 pA; $B_{ext}$ = 0.65 T). **d,** Energy-level diagram of two spins as a function of detuning $\omega_{RF} - \omega_0$. Avoided crossings are labeled as $\Delta_{S,-}$, $\Delta_{+,-}$ and $\Delta_{+,S}$. **e-g,** ESR spectra measured using a sinusoidal frequency modulation of $V_{RF}$, as a function of detuning $\omega_{RF} - \omega_0$ and frequency modulation amplitude $\delta_{RF}$ at a fixed modulation frequency $f$ of 20, 20, 30 MHz, respectively ($V_{bias}$ = 50

mV, $V_{RF}$ = 17, 20, 20 mV; $B_{ext}$ = 0.60 T; setpoint: $V_{sp}$ = 50 mV, $I_{sp}$ = 30, 30, 50 pA). The coupling $J$ is 0.2 GHz (**e**), 0.4 GHz (**f**) and 0.8 GHz (**g**). In **f**, white arrows indicate when the avoided crossings are reached for the static detuning indicated by the vertical dashed line. Insets show the STM images. **h,** Energy-level diagram of two spins as a function of the energy detuning $\delta_{bias}$ due to $V_{bias}$ near avoided crossings $\Delta_{S,-}$ (left) and $\Delta_{+,-}$ (right). **i-k,** ESR spectra measured under a square-wave amplitude-modulated $V_{bias}$, as a function of detuning $\omega_{RF} - \omega_0$ and modulation amplitude $\delta V$ at a fixed $f$ of 100 MHz ($V_{DC}$ = 50 mV, $V_{RF}$ = 15 mV; $B_{ext}$ = 0.59, 0.64, 0.66 T; setpoint: $V_{sp}$ = 50 mV, $I_{sp}$ = 60, 52, 70 pA). The white dashed lines indicate the boundaries between two branches of LZSM patterns.

In comparison, driving the spin dimer with amplitude-modulated $V_{bias}$ can only affect the energy detuning of $S_1$, which is different than using frequency-modulated $V_{RF}$ (Supplementary Sec. 4). Figure 5h shows the energy-level diagram as a function of energy detuning $\delta_{bias}$ due to $V_{bias}$ near avoided level crossings $\Delta_{S,-}$ and $\Delta_{+,-}$. For each RF frequency $\omega_{RF}$, we modulated the bias voltage $V_{bias}$, and the resulting LZSM interference patterns for spin dimers with different coupling strengths are plotted in Fig. 5i-k. At large coupling strength (Fig. 5k), the spectra manifest as a linear superposition of two single-spin interference patterns. However, as $J$ decreases, the LZSM patterns become more asymmetric, and the reduction in $J$ causes the two branches of interference patterns to move closer each other without intersecting (Fig. 5i,j). This behavior reflects repelling between the $|S\rangle$ and $|T_0\rangle$ states, which is absent under frequency modulation (Fig. 5e-g).

By combining RF frequency modulation (Fig. 5e-g) and the bias voltage modulation (Fig. 5i-k), we can thus explore different cross-sections of the energy-level diagram within the three-dimensional eigenenergy space (Fig. S3), defined by the total Zeeman energy of the two spins and the local Zeeman energy of the spin under the tip.

## Discussion

Our work demonstrates tunable LZSM interference at the atomic scale by electrical modulation of energy levels, spin-polarized current, and controlling the spin coupling via atomic manipulation. The all-electrical control of the spin dynamics near the avoided level crossings can be harnessed to realize fast coherent rotations of quantum states of surface spins and achieve robust quantum operations for quantum information processing[2,3,34]. Specifically, the LZSM protocol addresses the limitations of conventional resonant Rabi rotations[10,20] by leveraging quantum interference from non-adiabatic transitions, and exploits strong driving amplitudes using off-resonant harmonic signals[34,35]. In addition, the multi-level LZSM interference spectra demonstrated here can be applied to characterize the complex energy-level diagram of interacting spins on surfaces, which are complementary to the conventional single-atom ESR spectra[15,23,25,33]. Our electric control method can be readily applied to other atomic-scale magnetic structures, including spin chains[8], spin arrays[9,15,33], and molecular nanomagnets[36,37], providing new electric means for probing and manipulating quantum states at the atomic scale.

**Methods**

**Sample preparation.** Measurements were performed in a low-temperature STM (Unisoku USM1300) with home-built RF components for ESR measurement. MgO is two-monolayer thick, and was grown on Ag(001) single crystal by thermally evaporating Mg in an $\sim 10^{-6}$ Torr $O_2$ environment. Ti and Fe atoms were deposited from pure metal rods by e-beam evaporation onto the sample held at $\sim 10$ K. An external magnetic field of 0.51 T to 0.66 T was applied as indicated in the figure captions. STM images were acquired in constant-current mode, and all voltages refer to the sample voltage with respect to the tip.

**Spin-polarized tip.** The Pt-Ir STM tip was coated with silver by indentations into the Ag until the tip gave a good lateral resolution in the STM image. To prepare a spin-polarized tip, $\sim$1–5 Fe atoms were each transferred from the MgO onto the tip by applying a bias voltage ($\sim$0.55 V) while withdrawing the tip from near point contact with the Fe atom. The degree of spin polarization was verified by the asymmetry in $dI/dV$ spectra of Ti with respect to voltage polarity.

**Continuous-wave ESR measurement.** The continuous-wave ESR spectra were acquired by sweeping the frequency of an RF voltage $V_{RF}$ generated by the RF generator (Agilent E8257D) across the tunnel junction and monitoring changes in the tunnel current. The current signal was modulated at 95 Hz by chopping $V_{RF}$, which allowed readout of the current by a lock-in amplifier. The RF and bias voltages were combined at room temperature using an RF diplexer, and guided to the STM tip through semi-rigid coaxial cables with a loss of $\sim$30 dB at 20 GHz.

**Modulation of bias or RF voltages.** An arbitrary waveform generator (AWG, Tektronix AWG5204) is used to generate the amplitude-modulated bias voltage $V_{bias}$, which was transmitted to the tunnel junction via the RF diplexer. To generate the frequency-modulated RF voltage $V_{RF}$, two channels of the AWG were connected to the I and Q ports of an IQ mixer to modulate RF signal generated by the RF generator. The modulated RF signal was transmitted to the tunnel junction via the RF diplexer. See Extended Data Fig. 5 for the schematics of the experimental setup.

**Acknowledgments**

This work is supported by the National Natural Science Foundation of China (92476202 (K.Y., P.F.)), the Beijing Natural Science Foundation (Z230005 (K.Y., P.F.)), the National Key R&D Program of China (2022YFA1204100 (K.Y., L.J.)), the National Natural Science Foundation of China (12174433 (K.Y.), 52272170 (L.J.), 62488201 (H.-J.G.)), and the CAS Project for Young Scientists in Basic Research (YSBR-003 (K.Y.)). J.F-R. and Y.D.C acknowledge funding from FCT (Grant No. PTDC/FIS-MAC/2045/2021 and SFRH/BD/151311/2021). J.F-R. acknowledges funding from SNF Sinergia (Grant Pimag), the European Union (Grant FUNLAYERS-101079184), Generalitat Valenciana (Prometeo2021/017 and MFA/2022/045) and MICIN-Spain (Grants No. PID2019-109539GB-C41 and PRTR-C1y.I1). A.F. acknowledges ANPCyT (PICT2019-0654), CONICET (PIP11220200100170) and partial financial support from UNNE.

**Author contributions**

K.Y. designed the experiment. H.W., J.C., P.F., L.J. and K.Y. carried out the STM measurements. H.W., J.C., Y.D.C., A.F. and J.F-R. developed the theoretical model. H.W., J.C., Y.D.C., H.-J.G., H.F., and K.Y. performed the analysis and wrote the manuscript with help from all authors. All authors discussed the results and edited the manuscript.

**Competing interests**

The authors declare no competing interests.

**Data availability**

The data that support the plots within this paper and other findings of this study are available from the corresponding authors upon reasonable request.

**Extended data**

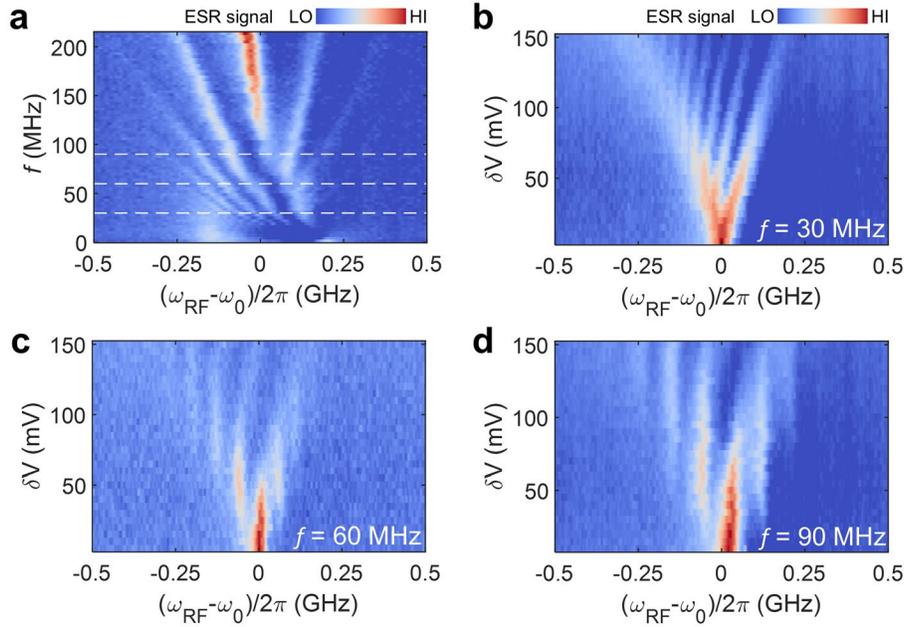

**Extended Data Fig. 1 | LZSM interference of a single Ti spin using bias voltage modulation. a,** ESR spectra as a function of detuning $\omega_{RF} - \omega_0$ and modulation frequency $f$ for a fixed modulation amplitude $\delta V$ of 100 mV ($V_{DC}$ = −50 mV, $V_{RF}$ = 15 mV; $B_{ext}$ = 0.59 T; setpoint: $V_{sp}$ = 50 mV, $I_{sp}$ = 30 pA). Dashed lines indicate the modulation frequencies at which **b-d** were taken. **b-d,** ESR spectra as a function of detuning $\omega_{RF} - \omega_0$ and modulation amplitude $\delta V$ for a fixed modulation frequency $f$ of 30, 60, 90 MHz, respectively ($V_{DC}$ = −50 mV, $V_{RF}$ = 20, 10, 20 mV; $B_{ext}$ = 0.59 T; setpoint: $V_{sp}$ = 50 mV, $I_{sp}$ = 30 pA).

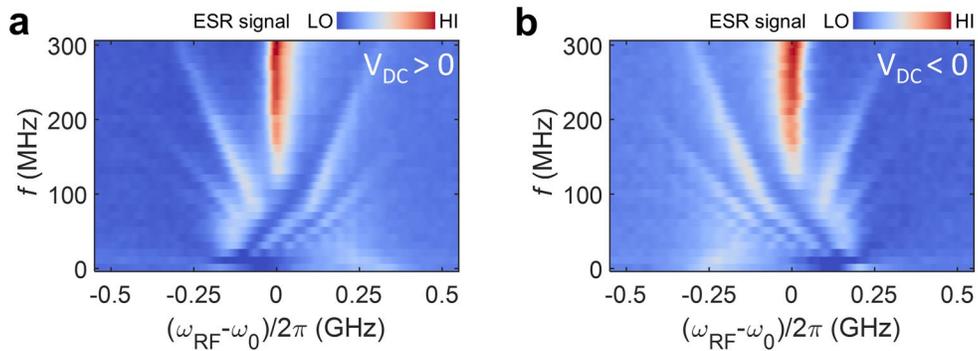

**Extended Data Fig. 2 | LZSM interference of a single Ti spin measured with positive and negative DC voltages. a,** ESR spectra as a function of detuning $\omega_{RF} - \omega_0$ and modulation frequency $f$ for a fixed modulation amplitude $\delta V$ ($V_{DC}$ = 50 mV, $\delta V$ = 100 mV, $V_{RF}$ = 15 mV; $B_{ext}$ = 0.51 T; setpoint: $V_{sp}$ = 50 mV, $I_{sp}$ = 60 pA). **b,** ESR spectra as a function of detuning $\omega_{RF} - \omega_0$ and modulation frequency $f$ for a fixed modulation amplitude $\delta V$ ($V_{DC}$ = −50 mV, $\delta V$ = 80 mV, $V_{RF}$ = 15 mV; $B_{ext}$ = 0.51 T; setpoint: $V_{sp}$ = 50 mV, $I_{sp}$ = 60 pA).

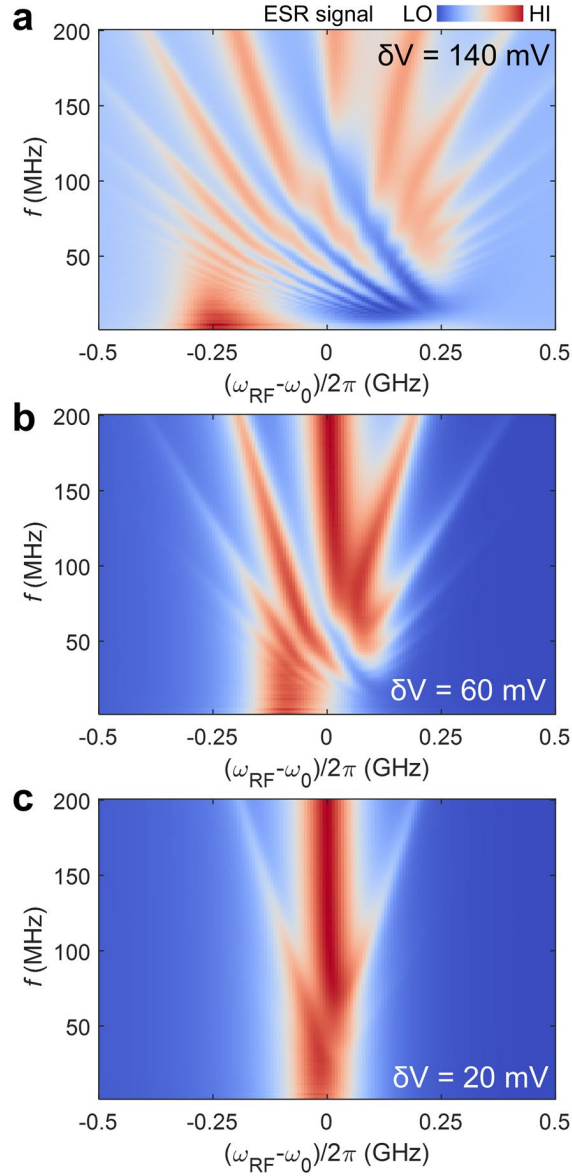

**Extended Data Fig. 3 | Simulations of LZSM interference by considering the spin-transfer torque effects using the generalized Bloch equations**. **a-c,** Simulated ESR signals as a function of detuning $\omega_{RF} - \omega_0$ and modulation frequency $f$ for a fixed modulation amplitude $\delta V$ of 140, 60 and 20 mV, respectively. Simulations parameters: $\omega_0$ = 15.5 GHz, $\Gamma_{STT}(t) = 11 - 9\sin(2\pi f t)$ MHz, $\Delta_{\uparrow\downarrow}$ = 60 MHz, $\delta_{bias}(t) = (40, 120, 280)\sin(2\pi f t)$ MHz, $V_{DC} = -50$ mV, $G_{junc}$ = 30 pS, $\alpha$ = 0.5, $\langle S_{tip}^z \rangle$=1, $\langle S_{tip}^{xy} \rangle$ =0.5, $V_{RF}$ = 30 mV, $T_1$ = 400 ns, $T_2$ = 200 ns.

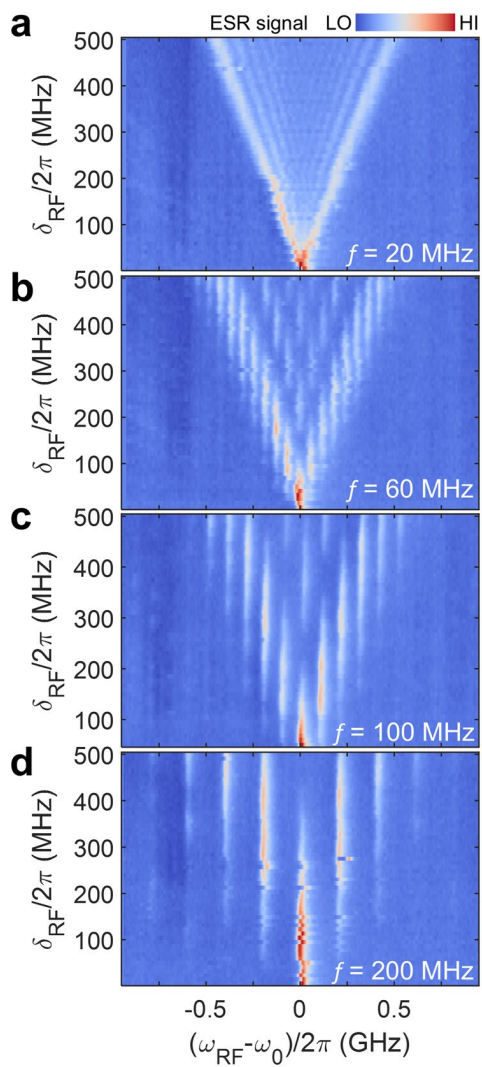

**Extended Data Fig. 4 | LZSM interference of a single Ti spin at different modulation frequencies using RF frequency modulation. a-d,** ESR spectra measured using a sinusoidal frequency modulation of $V_{RF}$, as a function of detuning $\omega_{RF} - \omega_0$ and frequency modulation amplitude $\delta_{RF}$ at a fixed modulation frequency $f$ of 20, 60, 100 and 200 MHz, respectively ($V_{bias}$ = 50 mV, $V_{RF}$ = 20 mV; $B_{ext}$ = 0.60 T; setpoint: $V_{sp}$ = 50 mV, $I_{sp}$ = 36 pA).

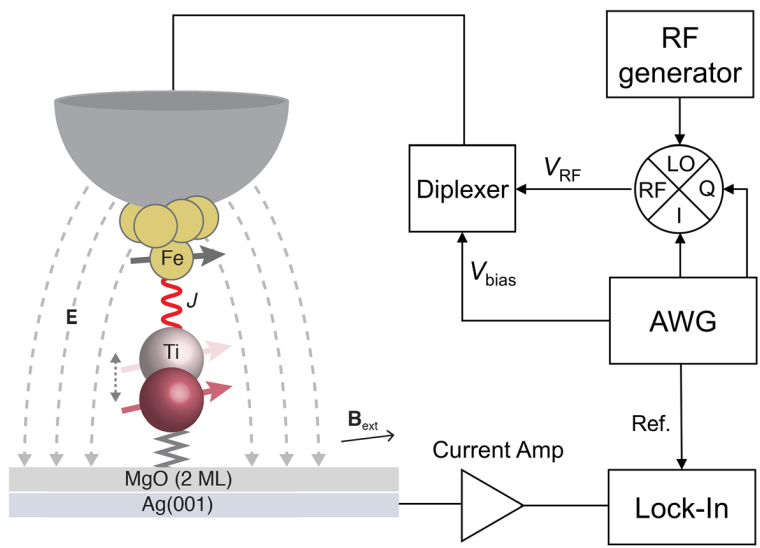

**Extended Data Fig. 5 | Schematics of the experimental setup for the measurement of LZSM interference.** Details of the setup are described in the methods section.